\documentclass[aps,prd,reprint,showpacs,preprintnumbers,superscriptaddress,nofootinbib]{revtex4-1}

\usepackage{slashed}
\usepackage{bm} % bold math
\usepackage{epsf}
\usepackage{graphicx,epsfig}
\usepackage{latexsym,amssymb,amsmath,float,url}
\usepackage{bbold}
\usepackage{latexsym}
\usepackage{soul}
\usepackage[normalem]{ulem} %Strikethrough using  \sout{text}.

\input{epsf}

\newcommand{\half}{\frac{1}{2}}

\providecommand \BibitemShut  [1]{\csname bibitem#1\endcsname}
\newcommand{\rarr}{\rightarrow}
\newcommand{\nubar}{\bar{\nu}}

\begin{document}

%%%%%%%%%%%%%%%%%%%%%%%%%%%%%%%%%%%%%%%%%%%%%%%%%
%%%%%%%%%%%%%%%%%%%%%%%%%%%%%%%%%%%%%%%%%%%%%%

%\preprint{}

\title{Dark Matter Annihilation Signatures from Electroweak Bremsstrahlung}

\author{Nicole F.\ Bell} 
%\email{n.bell@unimelb.edu.au}
\affiliation{School of Physics, The University of Melbourne, 
Victoria 3010, Australia}

\author{James B.\ Dent}
\affiliation{Department of Physics and School of Earth and Space Exploration,
Arizona State University, Tempe, AZ 85287-1404, USA}

\author{Thomas D.\ Jacques}
%\email{tjacques@ph.unimelb.edu.au}
\affiliation{School of Physics, The University of Melbourne, 
Victoria 3010, Australia}
\affiliation{Department of Physics and School of Earth and Space Exploration,
Arizona State University, Tempe, AZ 85287-1404, USA}

\author{Thomas J.\ Weiler}
%\email{t.weiler@vanderbilt.edu}
\affiliation{Department of Physics and Astronomy,
Vanderbilt University, Nashville, TN 37235, USA}

\date{23 May 2012}

\begin{abstract}
We examine observational signatures of dark matter (DM) annihilation in the
Milky Way arising from electroweak bremsstrahlung contributions to the
annihilation cross section.  It has been known for some time that
photon bremsstrahlung may significantly boost DM annihilation yields.
Recently, we have shown that electroweak bremsstrahlung of $W$ and $Z$
gauge bosons can be the dominant annihilation channel in some popular
models with helicity-suppressed $2\rarr 2$ annihilation.
$W$/$Z$-bremsstrahlung is particularly interesting because the gauge
bosons produced via annihilation subsequently decay to produce large
correlated fluxes of electrons, positrons, neutrinos, hadrons
(including antiprotons) and gamma rays, which are all of importance in
indirect DM searches.  Here we calculate the spectra of
stable annihilation products produced via
$\gamma/W$/$Z$-bremsstrahlung.  After modifying the fluxes to account
for the propagation through the Galaxy, we set upper bounds on the
annihilation cross section via a comparison with observational data.
We show that stringent cosmic ray antiproton limits preclude a sizable
DM contribution to observed cosmic ray positron fluxes in the
class of models 
for which the bremsstrahlung processes dominate.
\end{abstract}


% 95.35.+d  Dark matter
% 12.15.Lk  Electroweak radiative corrections
% 95.85.Ry  Neutrino, muon, pion, and other elementary particles; cosmic rays

\maketitle

%%%%%%%%%%%%%%%%%%%%%%%%%%%%%%%%%%%%%%%%%%%%%%%%
%%%%%%%%%%%%%%%%%%%%%%%%%%%%%%%%%%%%%%%%%%%%%%%%

%%%%%%%%%%%%%%%%%%%%%%%%%%%%%%%%%%%%%%%%%%%%%%%%
\section{Introduction}
\label{sec:intro}
%%%%%%%%%%%%%%%%%%%%%%%%%%%%%%%%%%%%%%%%%%%%%%%%

It has been firmly established that a significant fraction of the
energy density of the Universe resides in the form of dark matter
(DM), whose total abundance and approximate distribution has been
inferred via its gravitational
influence~\cite{Kamionkowski_review,Bertone_review,Bergstrom_review}.
However, we have as yet no direct detection, and the fundamental
particle properties of DM remain unknown.  A promising technique to
explore the particle nature of DM is to search for signatures of DM
annihilation or decay.  This is achieved by examining fluxes of
particles emanating from regions of DM concentration either in our own
Galaxy or throughout the cosmos.

Indirect DM detection has been the subject of much recent attention,
due to measured cosmic ray excesses 
of positrons and electrons above those expected from 
conventional astrophysical processes.  PAMELA has observed a sharp
excess in the $e^+/(e^- + e^+)$ fraction at energies beyond
approximately 10 GeV~\cite{PamelaPositrons}, without a corresponding
excess in the antiproton/proton data~\cite{PamelaAntiprotons,Adriani:2010rc},
while Fermi and HESS have reported more modest excesses in the $(e^- +
e^+)$ flux at energies of order 1 TeV~\cite{Fermi1,Aharonian:2009ah}.  
These signals
have led to a reexamination of positron production in nearby
pulsars~\cite{pulsars,Grasso:2009ma}, emission from supernova
remnants~\cite{supernova}, acceleration of $e^+e^-$ in cosmic ray
sources~\cite{accel}, and propagation in conventional cosmic ray
models~\cite{prop}.  Alternatively, it has been proposed the
excess electrons and positrons are not due to conventional
astrophysics process, but arise instead from DM annihilation
or decay in the Galactic halo.  A plethora of DM models have been
designed with this goal in mind (for a review see for example~\cite{dmmodels},
and references therein).

A viable resolution of the cosmic ray $e^\pm$ data by means of DM
annihilation requires a large branching ratio to leptons.  A large
branching ratio to hadrons would make a contribution to cosmic ray
antiproton fluxes, for which stringent observational bounds exist.
Therefore, so called leptophilic models are preferred, in which DM
couples (at tree level) only to leptons.  
However, for many scenarios in which the DM particle is a Majorana
fermion, annihilation to light fermions is helicity suppressed
($\propto m_f^2/s)$ in the $s$-wave contribution, and of course
velocity suppressed ($\propto v_{DM}^2$) in the $p$-wave contribution.
This is the case for popular DM candidates such as the neutralino of
supersymmetric models, if $\sqrt{s}\sim 2M_{DM}$ is below the $W^+W^-$
threshold; 
or Bino models with highly suppressed annihilation to $W^+ W^-$ and $ZZ$ final states.
 Large boost factors would be required for such a scenario
to explain any observed positron excess.

It has long been known that photon bremsstrahlung can lift helicity
suppressions~\cite{photonbrem,Bringmann:2007nk,Bergstrom:2008gr}.  We
have recently shown that helicity suppressions are also lifted by the
bremsstrahlung of a $W$ or $Z$ gauge boson~\cite{EWBrem2,EWBrem1} (see
also \cite{Ciafaloni:2011sa}).  (Electroweak radiative corrections to
DM annihilation, but without the observation of helicity
unsuppression, have been discussed in Ref.~\cite{Chen:1998dp,Berezinsky:2002hq,Kachelriess,BDJW,Dent:2008qy,Ciafaloni:2010qr,KSS09,Ciafaloni:2010b}.)
Where a helicity suppression is lifted, the cross sections for
$\chi\chi\rightarrow \ell\bar{\ell}\gamma$, $\chi\chi\rightarrow
\ell\bar{\ell}Z$ and $\chi\chi\rightarrow \ell\nu W$ can all greatly
exceed that for the lowest order process $\chi\chi\rightarrow
\ell\bar{\ell}$.  The bremsstrahlung processes thus may allow for the
indirect detection of many DM models which would otherwise be helicity
suppressed.

Importantly, the decay of the $W$ and $Z$ gauge bosons inevitably
leads to the production of secondary annihilation products, including
gamma rays, hadrons, charged leptons and neutrinos, allowing
multimessenger searches.  Note particularly, that even for DM models
designed to be leptophilic, production of hadrons is unavoidable.  (In
fact, even for models in which on-shell production of $W$ or $Z$ gauge
bosons is kinematically forbidden, some minimal hadron production is
inescapable, due to loop processes, or the exchange of off-shell
$W$ or $Z$ bosons.)

In this paper we examine a simple example of a model which has a
helicity-suppressed $2\rarr 2$ cross section, provided by
Ref.~\cite{Cao:2009yy,Ma:2000cc}.  We shall show that in this model,
the cross section required to produce positrons in sufficient quantity
to account for the observed excess will lead to overproduction of
antiprotons and gamma rays, and as such is ruled out as an explanation
of the observed positron anomalies.
Though our calculation is performed for the reference scenario of
Ref.~\cite{Cao:2009yy,Ma:2000cc}, we expect this conclusion to hold
for all scenarios in which the dominant positron contributions arise
from the 3-body $W$/$Z$-bremsstrahlung final states.

We calculate the spectra of both primary and secondary particles from
unsuppressed electroweak--bremsstrahlung annihilation processes, and
calculate the expected  spectra and fluxes
at Earth for a given annihilation cross
section.  We compare the Earthly fluxes with observational data to determine an
upper limit on the annihilation cross section.
While our analysis techniques are conservative, there are large
astrophysical uncertainties in the propagation of charged particles
through galactic magnetic fields, and in the DM density profile
which probably contains substructure.  A rigorous treatment of these
effects is beyond the scope of this work.  Consequently, our
constraints are illustrative of the upper limit on the cross section,
but not robust.

%%%%%%%%%%%%%%%%%%%%%%%%%%%%%%%%%%%%%
\section{Model}
\label{sec:model}
%%%%%%%%%%%%%%%%%%%%%%%%%%%%%%%%%%%%%

The example model we investigate is the Majorana DM version of the
leptophilic model proposed in~\cite{Cao:2009yy}.  Here the DM consists
of a gauge-singlet Majorana fermion $\chi$ which annihilates to
leptons via the interaction term
\begin{equation}
f\left(\nu\,\ell^-\right)_L\,\varepsilon\,
\left(
\begin{array}{l}
\eta^+ \\
\eta^0 \\
\end{array}
\right) \chi + h.c.
= f(\nu_L\eta^0 - \ell_L^- \eta^+)\chi + h.c.
\label{eq:ma}
\end{equation}
where $f$ is a coupling constant, $\varepsilon$ is the
$SU(2)$-invariant antisymmetric matrix, and $(\eta^+$, $\eta^0)$ form
the new $SU(2)$ doublet scalar which mediates the annihilation.  For
simplicity, we consider a coupling to the first generation of leptons
only, and set $f=0$ for coupling to the $(\nu_\mu\,\mu^-)_L$ and
$(\nu_\tau\, \tau^-)_L$ doublets.
As described in \cite{EWBrem1,EWBrem2}, the $p$-wave contribution to
the lowest order annihilation process $\chi\chi\rightarrow e^+e^-$ is
suppressed by $v_\chi^2\sim 10^{-6}$, while the $s$-wave contribution
is proportional to $(m_l/M_\chi)^2$.  This cross section is given by 
\begin{eqnarray}
v\,\sigma = \frac{f^4 v^2 }{24\pi\,M_\chi^2}\,\frac{1+\mu^2}{(1+\mu)^4}\,,
\label{eq:tree}
\end{eqnarray}
where $m_l\simeq0$ and $M_{\eta^\pm}=M_{\eta^0}$ have been assumed,
and $\mu=M_\eta ^2/M_\chi ^2$.
The helicity-suppressed $s$-wave term is absent in the $m_l=0$ limit,
leaving only the $v_\chi^2$-suppressed $p$-wave term.

While it is well known that photon bremsstrahlung $\chi\chi\rightarrow
e^+e^-\gamma$ can lift this suppression
\cite{photonbrem,Bergstrom:2008gr,Bringmann:2007nk},
Refs.~\cite{EWBrem2,EWBrem1} have shown that this is also the case for
the electroweak bremsstrahlung channels $\chi \chi \rightarrow e^+ e^-
Z, \,\nu_{e} \bar\nu_{e} Z, \,e^+ \nu_{e} W^-, \,e^- \bar\nu_{e} W^+$.  For both
$W$/$Z$ and $\gamma$ bremsstrahlung, the effect is most significant
where the DM mass is nearly degenerate with the mass of the
boson which mediates the annihilation process.
In the high energy limit where the $W$/$Z$ masses are negligible, the
cross sections for $W$/$Z$ bremsstrahlung reduce to that for $\gamma$
bremsstrahlung, modulo different coupling constants.  However, the
respective sizes of the electromagnetic and electroweak coupling
constants imply that the $W$/$Z$-strahlung cross section is a factor
of several larger than that for $\gamma$-strahlung,
\begin{eqnarray}
\sigma_{e^+\nu_{e} W^-} &=& \sigma_{e^-\bar{\nu}_{e} W^+} = \frac{1}{2 \sin^2\theta_W}  \, \sigma_{e^+e^-\gamma},\\
\sigma_{\bar\nu_{e} \nu_{e} Z} &=&  \frac{1}{4 \cos^2\theta_W \sin^2\theta_W}  \, \sigma_{e^+e^-\gamma},\\
\sigma_{e^+e^- Z} &=& \frac{\left(\frac{1}{2}-\sin^2\theta_W \right)^2}{\cos^2\theta_W\sin^2\theta_W}  \, \sigma_{e^+e^-\gamma},
\end{eqnarray}
and thus 
\begin{eqnarray}
v\,\sigma_{W/Z-\textrm{strahlung}} &=& 6.16 \,v\, \sigma_{e^+e^-\gamma}. 
\end{eqnarray}
At lower energies, the phase space for $W$/$Z$ bremsstrahlung is
somewhat reduced due to the effects of the finite $W$/$Z$ masses.

The bremsstrahlung cross section dominates over that for the lowest
order $2\rarr 2$ process, provided that $M_\eta$ does not greatly
exceed $M_\chi$.  The ratio $R_W=v \sigma_{e^+ \nu_{e} W^-}/ (v
\sigma_{e^+ e^-})$ was plotted in Ref.~\cite{EWBrem2}, and is largest
for $\mu = (M_\eta/M_\chi)^2=1$.  However, the $W$-strahlung process
dominates over the $2\rarr 2$ annihilation even if a mild hierarchy
between $M_\chi$ and $M_\eta$ is assumed, with $R_W>1$ for $\mu
\lesssim 10$.  Therefore, when $\mu \lesssim 10$, production of
monoenergetic leptons will be subdominant to particles produced in
the 3-body processes (both primary and through gauge boson decay), a
feature which must be accounted for when analysing astrophysical
signatures of these models.

We use the cross sections calculated in
\cite{Bergstrom:2008gr,Bringmann:2007nk,EWBrem2} in combination with
the PYTHIA code~\cite{pythia1,pythia2} to determine the spectra of
gamma rays, electrons, protons, and their antiparticles, per
annihilation to the five 3-body final states listed above (one
electromagnetic bremsstrahlung and four electroweak bremsstrahlung
processes).  After accounting for propagation effects for the charged
particles, we constrain these cross sections by comparing the observed
flux with the calculated annihilation signal.

Although the spectra of annihilation products which we show are unique
to the particular model we have chosen, we expect the results to apply
qualitatively to any model where $W/Z$-bremsstrahlung is the dominant
annihilation mode (i.e.\ where helicity suppression of the $2\rarr 2$
$s$-wave is lifted by electroweak bremsstrahlung).

%%%%%%%%%%%%%%%%%%%%%%%%%%%%%%%%%%%%%
\section{Cross Section Channel}
\label{sec:channel}
%%%%%%%%%%%%%%%%%%%%%%%%%%%%%%%%%%%%%

As mentioned, the model we examine is leptophilic in the
$2\rarr 2$ process, and therefore has five 3-body bremsstrahlung
channels, $\chi \chi \rightarrow e^+ e^- Z, \nu_{e} \bar\nu_{e} Z, e^+ \nu_{e}
W^-, e^- \bar\nu_{e} W^+, e^+e^-\gamma$, which simultaneously contribute
to the DM annihilation fluxes.  The cross sections and
spectra for the electroweak channels are specified in
Ref.~\cite{EWBrem2}, while those for the electromagnetic channel are
given in Refs.~\cite{Bergstrom:2008gr,Bringmann:2007nk}.  The total
bremsstrahlung cross section is given by the sum
\begin{eqnarray}
v \sigma_{\rm Brem}&=&v \sigma_{e^+ e^- Z} + v \sigma_{\nu_{e} \bar\nu_{e} Z}
\nonumber\\&+&
v \sigma_{e^+ \nu_{e} W^-}+v \sigma_{ e^- \bar\nu_{e} W^+} + v \sigma_{e^+ e^- \gamma}.
\end{eqnarray}
The total bremsstrahlung cross section is a factor of $\sim 7.2$
larger than that for photon bremsstrahlung alone, due to the four $W$/$Z$
channels, which are governed by somewhat larger coupling constants.

We shall consider parameters for which the bremsstrahlung channels
dominate the total cross section, so that $v \sigma_{\rm Brem} \simeq
v \sigma_{total}$.
We define the ``branching ratio'' for an individual channel
$i \in \{e^+ e^- Z,\nu_{e} \bar\nu_{e} Z, \,e^+ \nu_{e} W^-, \,e^- \bar\nu_{e} W^+, \,e^+ e^-\gamma\}$ as 
\begin{equation}\label{Brbrem}
BR_{\rm Brem}(i)=\dfrac{v \sigma_i}{v \sigma_{\rm Brem}}.
\end{equation}
The spectrum per annihilation, for any given annihilation product, 
$k \in \{\gamma, \,e^-, \,e^+, \,\nu, \,\bar\nu, \,p, \,\bar p \}$, is then given by
\begin{equation}
\left.\frac{dN_k}{dE_k}\right|_{\rm Brem}
= \sum_i BR_{\rm Brem}(i) \left.\frac{dN_k}{dE_k}\right|_{{\rm per } \,\chi\chi\rightarrow i}.
\end{equation}
Here, $\left.\frac{dN_k}{dE_k}\right|_{{\rm per } \,\chi\chi\rightarrow
  i}$ is the spectrum per annihilation for a given channel.  
  The spectra for $\gamma,e^\pm,\ \nu$ and $\bar{\nu}$  include primary annihilation products 
  and secondary annihilation products produced by gauge boson fragmentation.
  The spectra for $p$ and $\bar{p}$ arise exclusively from gauge boson fragmentation.
The branching ratios and spectra depend on the parameter
$\mu=(M_\eta/M_\chi)^2$.  However, as long as the 3-body final states
remain the dominant channel, the spectra (and thus the final results)
have little dependence on this parameter.  We show results for
$\mu=(M_\eta/M_\chi)^2 = 1.2$, but results remain qualitatively
unchanged 
when bremsstrahlung channels dominate over $2\rarr 2$ processes.

Finally, the flux of a given annihilation product is schematically 
\begin{equation}
\frac{d\phi_k}{dE_k}
\propto v \sigma_{\rm Brem} \left.\frac{dN_k}{dE_k}\right|_{\rm Brem}.
\end{equation}
The detailed evaluation of these annihilation spectra is given below.

%%%%%%%%%%%%%%%%%%%%%%%%%%%%%%%%%%%%%%%%%%%%%%%%
\section{Annihilation Spectra}
\label{sec:spectra}
%%%%%%%%%%%%%%%%%%%%%%%%%%%%%%%%%%%%%%%%%%%%%%%%

\noindent

In order to place constraints on the cross section, we need the spectrum 
of stable particles ($\nu, e^-, p$ and their antiparticles,
plus $\gamma$) produced per DM annihilation. 
As an example, we describe how we determine the spectrum of antiprotons per
$\chi\chi\rightarrow \nu_{e} e^+W^-$ event; the technique is very similar
for other secondary particles, and other electroweak--bremsstrahlung
annihilation channels.
These partial spectra are then summed to form $\left.\dfrac{dN_k}{dE_k}\right|_{\rm Brem}$.

We use the PYTHIA code \cite{pythia1,pythia2} to find the spectrum of
antiprotons per $W^-$ decay, $\left.dN_{\bar p}/dE\right|_{\rm W
  decay}$.  We produce a $W^-$ boson in its rest frame by colliding an
antimuon with a muon neutrino, with center of mass energy $M_W$, and
turning off all processes other than $\mu^\pm\nu_\mu
(\nubar_\mu)\rightarrow W^\pm$.  (Similarly, to produce the $Z$ boson,
we collide a $e^+e^-$ pair at CoM energy $M_Z$, leaving $Z$ production
as the only active process.)  Unstable $W^-$ decay products
(mainly pions) themselves decay, finally  leaving only
neutrinos, electrons, protons and their antiparticles, plus gamma
rays in the final state.  These stable particles are placed 
in 2000 logarithmically-spaced energy bins.  
The final spectrum is found by averaging the PYTHIA spectra over 10,000 such events.

We use this energy spectrum, in combination with the $W^-$ (or $Z$)
energy distribution per annihilation, $dN_W/d\gamma=(1/v \sigma)(dv
\sigma /d\gamma)$, where $\gamma=E_W/M_W$, to find the antiproton
energy-spectrum per annihilation in the lab frame (see
Appendix of~\cite{EWBrem1} for a derivation)
\begin{equation}\label{E1}
\frac{dN_{\bar p} (E')}{dE'}=\half\int_1^\infty \frac{d\gamma}{\sqrt{\gamma^2-1}}\frac{dN_W}{d\gamma}
     \int_{E_-}^{E_+} \frac{dE}{p}\,\frac{dN_{\bar p}}{dE}\,,
\end{equation}
with $p=\sqrt{E^2-m^2_{\bar p}}$, $\beta\gamma=\sqrt{\gamma^2-1}$, and
$E_\pm = \gamma E'\pm\beta \gamma p'$.  Or equivalently, 
\begin{equation}\label{E2}
\frac{dN_{\bar p} (E')}{dE'}=\half\int_{m_{\bar p}}^\infty \frac{dE}{p}\,
     \frac{dN_{\bar p}}{dE}\,\int_{\gamma_-}^{\gamma_+} 
     \frac{d\gamma}{\sqrt{\gamma^2-1}}\frac{dN_W}{d\gamma}\,,
\end{equation}
with $\gamma_\pm=(EE'\pm pp')/m^2_p$ and $p'=\sqrt{E'^2-m^2_p}$.
In the case of the gamma-ray, electron, positron, and neutrino
spectra, we add the spectrum of primary annihilation products to the
spectrum of secondaries from $W$~decay, calculated as above.

\begin{figure}
\includegraphics[width=0.45\textwidth]{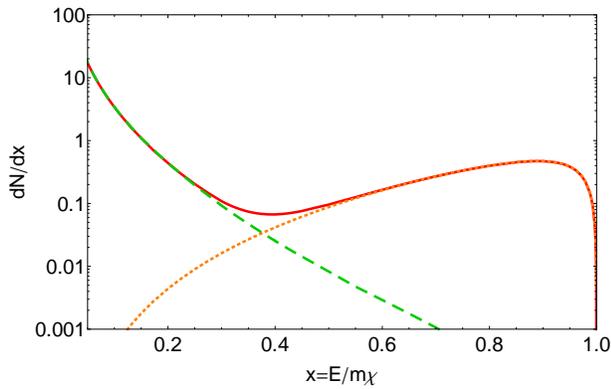}
\caption{Contributions to the gamma-ray spectrum per annihilation,
  $\left.\dfrac{dN_{\gamma}}{dE}\right|_{\rm brem}$, from primary
  production in photon bremsstrahlung (dotted, orange), and $W$/$Z$
  decay products (dashed, green), for $M_\chi=300$ GeV and $M^2_\eta/
  M^2_\chi=1.2$.  The total gamma-ray spectrum is shown as a solid
  curve (red).
\label{fig:contributions-gamma}}
\end{figure}

\begin{figure}
\includegraphics[width=0.45\textwidth]{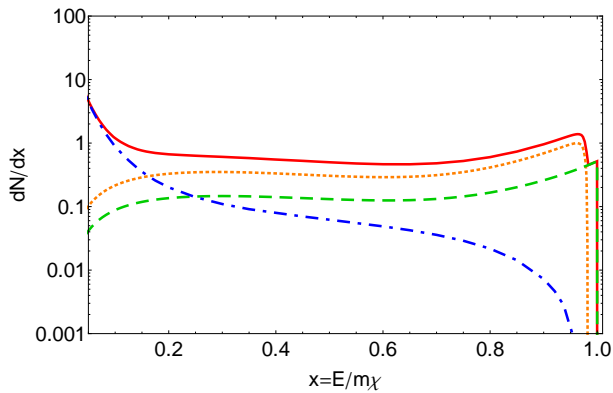}
\caption{Contributions to the positron spectrum per annihilation,
  $\left.\dfrac{dN_{e^+}}{dE}\right|_{\rm brem}$, from primary
  production in electroweak bremsstrahlung channels (dotted, orange),
  primary production in the photon bremsstrahlung channel (dashed,
  green) and $W$/$Z$ decay products (dot-dashed, blue), for
  $M_\chi=300$ GeV and $(M_\eta/ M_\chi)^2=1.2$.  The total positron
  spectrum is shown as a solid curve (red). Note that the positron
  spectra from electroweak and photon bremsstrahlung have differing
  kinematic cutoffs due to the masses of the $W^\pm$ and $Z$ bosons.
\label{fig:contributions-e}}
\end{figure}

\begin{figure}
\includegraphics[width=0.45\textwidth]{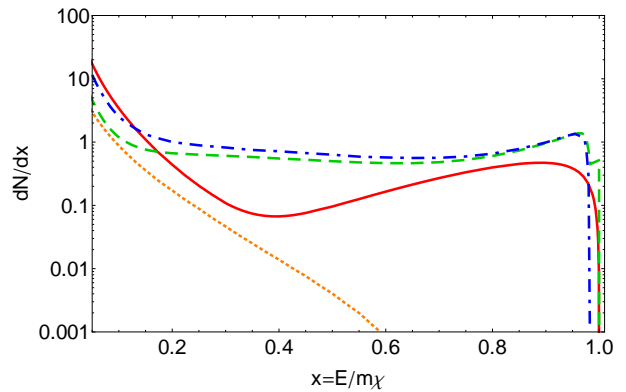}
\caption{Spectrum per annihilation of photons (solid, red), protons
  (dotted, orange), electrons (dashed, green) and neutrinos
  (dot-dashed, blue), for $M_\chi=300$ GeV and $M^2_\eta/
  M^2_\chi=1.2$. For protons, $E$ is the kinetic energy.  By
  CP--invariance, the particle and antiparticle spectra are the same,
  and antiparticles are not included in this figure.  Note that the
  neutrino spectrum includes primary electron neutrinos, and all
  flavors of secondary neutrinos.
\label{fig:spectra}}
\end{figure}

Figure~\ref{fig:contributions-gamma} shows the total gamma-ray
spectrum, as well as the relative contributions from primary and
secondary annihilation products, clearly showing that the secondary
gamma rays are subdominant, except at low energy.
Figure~\ref{fig:contributions-e} shows the same information for the
positron spectra, including the relative contributions to the primary
positron spectrum from photon bremsstrahlung and electroweak
bremsstrahlung, as well as the secondary positrons produced via gauge
boson decay.
In Fig.~ \ref{fig:spectra}, we show the total spectra per
annihilation for electrons, neutrinos, protons, and gamma rays.
Note that the electron/positron spectra from $W$/$Z$ and photon
bremsstrahlung have differing kinematic cutoffs due to the masses of
the $W^\pm$ and $Z$ bosons, leading to a kink near the endpoint in the
electron/positron spectra.  This feature is absent from the neutrino
spectrum, as there is no contribution from photon bremsstrahlung.
The neutrino spectrum includes contributions from primary electron neutrinos,
and all flavors of secondary neutrinos from $W^\pm/Z$ decay.
(The flavor ratios of primary neutrino production are model-dependent, given by 
$f^2_e : f^2_\mu : f^2_\tau$. We have assumed $f_{\mu} = f_{\tau} = 0$.)

%%%%%%%%%%%%%%%%%%%%%%
%%%%%%%%%%%%%%%%%%%%%%

%%%%%%%%%%%%%%%%%%%%%%
%%%%%%%%%%%%%%%%%%%%%%

%%%%%%%%%%%%%%%%%%%%%%
%%%%%%%%%%%%%%%%%%%%%%

It is illuminating to compare our spectra to those for annihilation to
a pair of gauge bosons.  The photon spectrum for the WIMP annihilation
channel $\chi\chi \rightarrow W^+W^-$, shown for example in Fig.~1 of
Cembranos et al.~\cite{Dobado}, has a somewhat softer gamma spectrum,
while we have a somewhat harder spectrum with more higher-energy
photons.  This is to be expected, as gamma rays in Ref.~\cite{Dobado} arise 
only from decay of monoenergetic $W$ bosons ($E_W = M_\chi$), while
the photon-bremsstrahlung process contributes a harder primary gamma-ray spectrum.  
The reverse holds true for the proton spectra, as our
3-body $W$/$Z$-bremsstrahlung process results in a broad distribution
of $W$ energies (the spectrum is shown in Fig.~3 of
Ref.~\cite{EWBrem2}).  In addition, the electron and neutrino spectra
resulting from the $\chi\chi \rightarrow W^+W^-$ process would be
quite different to those for electroweak bremsstrahlung, given that
for the latter it is the primary leptons (not the secondaries from
$W$/$Z$ fragmentation) that make the dominant contribution.
Of course, the spectra of charged particles observed at Earth will
differ from that at production, due to the effect of energy loss
processes during propagation.  We address these effects in the
following section.

%%%%%%%%%%%%%%%%%%%%%%%%%%%%%%%%%%%%%
\section{Constraints}
\label{sec:constraints}
%%%%%%%%%%%%%%%%%%%%%%%%%%%%%%%%%%%%%

In this section, we place conservative upper limits on the
thermally-averaged self-annihilation cross section, $\langle v
\sigma\rangle_{\rm Brem} \simeq \langle v \sigma\rangle_{total}$, for
the leptophilic model described above.
We do this by following the same technique as in, for example,
\cite{bj,MJBBY,KSS09}. We compare the various predicted fluxes for a
particular DM annihilation channel with the relevant
observational flux measurements.  We make the conservative assumption
that the entire observed flux comes from DM annihilation; in reality,
astrophysical backgrounds are likely to contribute a large fraction of
the observed fluxes.  The upper limit on the cross section is then
determined such that the DM annihilation does not exceed any of the
observed fluxes.

In calculating the constraints on $\langle v \sigma \rangle _{\rm
  Brem}$, we utilize the isotropic extragalactic gamma-ray flux
measured by the Fermi collaboration \cite{FermiEGB}, the very-high-energy gamma-ray 
flux from H.E.S.S. \cite{Abramowski:2011hc}, the positron
fraction from the PAMELA collaboration \cite{PamelaPositrons}, the
Fermi $e^++e^-$ flux \cite{Fermi1}, as well as the antiproton flux
and antiproton-to-proton ratios updated by the PAMELA
collaboration in Ref.~\cite{Adriani:2010rc}.
Wherever uncertainties in the flux are presented, we use the 1-$\sigma$ upper limit.
Throughout, we use the commonly-adopted Navarro, Frenk and White (NFW)
DM density profile \cite{NFW}, with local DM density
given by $\rho_\odot = 0.39 \textrm{ GeV /cm}^{-3}$
\cite{Catena:2009mf}.\footnote { $\rho_\odot = 0.389 \pm 0.025$
  (1-$\sigma$) \cite{Catena:2009mf}, but could be a factor of $\sim 2$
  higher or lower \cite{Weber:2009pt,Salucci:2010qr,Cholis:2010}.}
We properly account for proton and electron diffusion and energy loss as detailed below.

%%%%%%%%%%%%%%%%%%%%%%%%%%%%%%%%%%%%%
\subsection{Gamma Rays}
\label{subsec:gammas}
%%%%%%%%%%%%%%%%%%%%%%%%%%%%%%%%%%%%%

The isotropic diffuse gamma-ray flux will have contributions from both
galactic and extragalactic DM annihilation. Although the
galactic signal is expected to have a large directional dependence,
there will be an underlying isotropic component \cite{yhba}.
We include both contributions when computing constraints, though the
galactic flux dominates over the extragalactic for the parameters of
interest.
In order to calculate the cosmic annihilation signal, we follow the technique set out in Refs. \cite{Ullio:2002pj,Bergstrom:2001jj}, while adopting the notation of \cite{MJBBY,yhba}.

The isotropic gamma-ray flux from DM annihilations throughout the Universe is given by
\begin{eqnarray}
\frac{d\Phi_\gamma}{dE} &=& \frac{\langle v \sigma \rangle}{2}
\frac{c}{4 \pi H_0} \frac{\rho^2_{\textrm{av}}}{M_\chi^2}\nonumber \\
&\times& {\int_0}^{z_{up}} \frac{f(z)(1+z)^3}{h(z)}
\frac{dN_\gamma(E')}{dE'} e^{-\tau(z,E)}dz\,,
\label{cosmicflux}
\end{eqnarray}
where $H_0 = 70$ km s$^{-1}$ Mpc$^{-1}$ is the Hubble parameter,
$\rho_{\textrm{av}}$ is the average DM density in the
Universe, and $h(z) = [(1+z)^3 \Omega_{DM} +\Omega_\Lambda]^{1/2}$. We
assume $\Omega_{DM}$ = 0.3, $\Omega_\Lambda$ = 0.7.  The energy at
production is $E'$, and the redshifted energy is $E=E'/(1+z)$.  We use
the optical depth $\tau(z,E)$ from \cite{Ullio:2002pj}, which accounts
for attenuation of gamma rays as they propagate through the Universe.
The factor $f(z)$ accounts for the clustering of DM, which
gives an enhancement of the annihilation signal relative to a universe
in which matter was distributed homogeneously.  Following~\cite{yhba},
we parametrize the redshift dependence as $\log_{10}(f(z)/f_0) = 0.9
\,[\textrm{exp}(-0.9z)-1]-0.16z$.  There is some debate as to the
overall normalization of $f(z)$.  We adopt the NFW profile, which
gives $f_0 \simeq 5 \,\times 10^{4}$.  This is a conservative choice
for $f(z)$, and inclusion of enhancements due to subhalos would only
strengthen our results.  See Ref.~\cite{Abdo:2010dk} for a recent
discussion of the clustering factor.  Our choice for $f(z)$ closely
corresponds to one of the smallest examples given there.

For Galactic annihilation, we again follow the technique of
e.g.~\cite{yhba,MJBBY}.  The flux of gamma rays per steradian from
Galactic DM annihilation, in a direction at an angle $\psi$ from the
Galactic Center, is given by
\begin{eqnarray}
\frac{d\Phi_{\gamma}}{dE} &=& \frac{1}{2}\frac{\langle v\sigma \rangle}{4 \pi M_\chi ^2} \frac{\mathcal{J}(\psi)}{{\rm J}_0}\frac{dN_\gamma}{dE}\label{gamma-flux1},
\end{eqnarray}
where
\begin{equation}
{\cal J}(\psi) = {\rm J_0}
\int^{\ell_{max}}_0 \rho^2\left(\sqrt{R_{\textrm{sc}}^2 -
2\ell R_{\textrm{sc}}\cos{\psi} +\ell^2}\right)d\ell 
\label{los}
\end{equation}
is the integral along the line of sight of the DM density
squared.  J$_0=1/[8.5\,{\rm kpc} \times(0.3 \,{\rm GeV \,
    cm}^{-3})^2]$ is an arbitrary normalization constant used to make
$\cal{J}(\psi)$ dimensionless; it cancels from our final expression
for the gamma-ray flux.  Since we are calculating the isotropic
signal, we use the minimum value for ${\cal J}(\psi)$, once again
adopting the NFW profile.

The Extragalactic Gamma-Ray Background (EGB) reported by Fermi in
\cite{FermiEGB} is the isotropic component of the diffuse gamma-ray
flux, with a number of potential contributing sources. It is obtained
by subtracting the components of the gamma-ray flux with known origin
from the total flux, observed away from the Galactic disk (Galactic
latitude $|b| \geq 10^\circ$).  Hence, it is a flux likely to contain
a contribution from either Galactic or extragalactic DM annihilation.
We compare our calculated isotropic signal, from both cosmic and
Galactic annihilation, to this isotropic flux.  We do this for each
data energy bin, integrating the signal over the width of each bin in
turn.

We also compare the DM annihilation signal with H.E.S.S. observations
of the very-high-energy gamma-ray flux from an angular region of 1$^\circ$
 around the Galactic Center, excluding the Galactic Plane \cite{Abramowski:2011hc}. 
 These observations have both advantages and disadvantages  when
 compared with the Fermi EGB observations. First, the two data sets cover different energy regimes. The H.E.S.S. data
 begin at around 300 GeV, and so are unable to constrain the cross section at lower DM masses.
 Conversely, Fermi EGB data end at 100 GeV, such that constraints on the cross section
 for larger DM masses are based on comparisons of the observed flux with the tail-end of the annihilation spectrum, leading to relatively weak constraints.  
In addition, the two data sets are from drastically different observation regions.
The expected isotropic DM signal is relatively small, leading to weaker constraints from the Fermi data; however, this is balanced by the small uncertainty between the various DM density profiles for this region. The opposite holds true for the H.E.S.S. observation region, where there is a large expected DM signal, large background flux, and a very large uncertainty between competing density profiles, since the observation region is small and close to the Galactic Center. While we report results using the NFW profile, these would be weakened by well over an order of magnitude if an isothermal profile were adopted.

For these reasons it makes sense to constrain the annihilation rate using both data sets, emphasizing that while we find that the H.E.S.S. constraints are significantly stronger, they are subject to larger uncertainties. We follow the same technique as earlier, treating the flux from the Source region in Ref.~\cite{Abramowski:2011hc} as an upper limit on the annihilation flux calculated using Eq.~\ref{gamma-flux1} for each reported energy bin, allowing us to place an upper limit on $\langle v \sigma\rangle$ as a function of the DM mass. 
We again present results using the NFW profile, utilizing the average of $\mathcal{J}(\psi)$ over the observation region as reported in Ref.~\cite{Abramowski:2011hc}, $\mathcal{J}_{\rm av} = 1604$.

Our resulting upper limits on $\langle v \sigma\rangle$ using both  Fermi EGB and H.E.S.S. data
are reported in Fig.~\ref{fig:limits-brem}.

%%%%%%%%%%%%%%%%%%%%%%%%%%%%%%%%%%%%%
\subsection{Electrons and Positrons}
\label{subsec:positrons}
%%%%%%%%%%%%%%%%%%%%%%%%%%%%%%%%%%%%%

The flux of positrons (or electrons) at Earth from DM
annihilation depends both on the propagation of the positrons through
the turbulent galactic magnetic fields, and energy losses of the
particles.  
One can solve the applicable
diffusion-energy loss equation to find a semianalytic form for the
positron flux at Earth. We adopt the same notation and set of assumptions 
as in e.g. Ref~\cite{Cirelli:2008id}, which gives
\begin{equation}
\frac{d\Phi_e(E)}{dE}= \frac{\langle v \sigma \rangle\, \rho_\odot^2 v}{8 \pi M_\chi^2 b(E)}\int_E^{M_\chi}dE'
\frac{dN_e}{dE}I\left(\lambda_D\left(E,E'\right)\right),
\end{equation}
where $\rho_\odot$ is the local DM density, $b(E)$ and
$I\left(\lambda_D\left(E,E'\right)\right)$ are the energy loss and
``halo function'' parameters respectively, and
$\lambda_D\left(E,E'\right)$ is the diffusion length between the two
energies $E$ and $E'$.  The energy loss of the positrons as they propagate
through the Galactic medium is mainly due to synchrotron radiation and
inverse Compton scattering, and is characterized by $b(E)\approx 10^{-16}
(E/{\rm GeV})^2$ GeV s$^{-1}$.
The ``halo function'' $I\left(\lambda_D\left(E,E'\right)\right)$ is an
astrophysical parametrization which encodes the dependence of the flux
on the DM density profile, and on the model of positron
diffusion due to Galactic magnetic fields.  There is a degree of
uncertainty in this function, and so we choose the `medium'
diffusion parameter set (while also showing results in the  `min' and `max' scenarios), 
and as usual, the NFW DM density profile.
We use the numerical fit to
$I\left(\lambda_D\left(E,E'\right)\right)$ from~\cite{Cirelli:2008id}.

Our signal is then compared with the total $e^++e^-$ flux reported by 
the Fermi collaboration~\cite{Fermi1} to find an upper limit on $\langle v \sigma\rangle$, 
by demanding that the signal integrated over the width of an energy bin be less than the total observed flux in that bin,
\begin{equation}
(\Phi_{e^+}+\Phi_{e^-})^{\rm signal} = 2\,\Phi_{e^+}^{\rm signal} \lesssim \Phi_{e^++e^-}^{\rm obs}.
\end{equation}
We can also combine our positron flux with the Fermi data to find the positron fraction from DM annihilation. We compare this with the PAMELA data~\cite{PamelaPositrons} for the 
positron fraction $(f_{e^+})$ to find an alternative upper limit on $\langle v \sigma \rangle$ by demanding
\begin{equation}
\frac{\Phi_{e^+}}{\Phi^{\rm obs}_{e^++e^-}}\leq f_{e^+}\,.
\end{equation}
We compare the DM-related positron
fraction with the observed PAMELA fraction in each of the four energy
bins where the Fermi energy range overlaps the PAMELA energy range,
integrating the DM-signal and observed Fermi fluxes over the width
of the PAMELA energy bins.
For this, we use the simple power-law fit to the Fermi data, valid between around 20 GeV and 1 TeV \cite{Grasso:2009ma},
\begin{eqnarray}
\frac{d\Phi^{\rm obs}_{e^++e^-}}{dE}&=&(175.40\pm6.09)\times 10^{-4}\,({\rm GeV\,cm}^{2}\,{\rm s\,sr})^{-1}
\nonumber\\&\times &
 (E/{\rm GeV})^{(-3.045\pm 0.008)}\,.
\end{eqnarray}
Results are reported in Fig.~\ref{fig:limits-brem}, 
 while results in the `minimum' and `maximum' diffusion scenarios are shown in Fig.~\ref{fig:limits-comparison}.

%%%%%%%%%%%%%%%%%%%%%%%%%%%%%%%%%%%%%
\subsection{Protons and Antiprotons}
\label{sec:protons}
%%%%%%%%%%%%%%%%%%%%%%%%%%%%%%%%%%%%%

The antiproton (or proton) flux at Earth has a similar functional form
to that for positrons, except that the energy losses for the
antiprotons as they propagate to Earth are negligible, since $m_p \gg
m_e$.  (Note that although energy loss is negligible, diffusion is
not.)  Because the energy for the antiprotons is the same as the
injection energy, there is no need for an integration over energies $E'$ at
production.
(Energy loss due to scattering
interactions or solar modulation are relevant only at low energies.)
We again use the semianalytic function from \cite{Cirelli:2008id} to
calculate the proton and antiproton signals at Earth from DM
annihilation,
\begin{equation}
\frac{d\Phi_p(K)}{dK}=\frac{\langle v \sigma \rangle\, \rho_\odot^2 v}{8 \pi M_\chi^2} R(K)
\end{equation}
where $K$ is the kinetic energy of the (anti)-proton, and $R(K)$ is an
astrophysics parametrization playing a similar role to
$I\left(\lambda_D\left(E,E'\right)\right)$ from Sec.~\ref{subsec:positrons}.  Reference~\cite{Cirelli:2008id} provides a
numerical fit to $R(K)$ for several sets of propagation parameters,
and we again use the `medium' parameter set, with the NFW DM density
profile. 

We compare our antiproton flux with the total antiproton flux reported by the PAMELA collaboration~\cite{Adriani:2010rc},
energy bin by energy bin.
We can also constrain the cross section by demanding 
the ratio $\bar p/p$ due to antiprotons from DM annihilation 
not exceed the PAMELA $\bar p/p$ ratio \cite{PamelaAntiprotons,Adriani:2010rc}.
This comparison requires the observed proton flux.
Following \cite{KSS09}, we use the nucleon flux from \cite{PDG}, 
$\frac{d\Phi_p^{\rm obs}}{dE}\approx 0.79\times 1.8 (E/{\rm GeV})^{-2.7}$~(cm$^2$\,s\,GeV)$^{-1}$,
where $0.79$ is the proton fraction of the total nucleon flux.
We then simply demand $\Phi_{\bar p}/\Phi_{p}^{\rm obs} \lesssim f_{\bar p/p}$, 
where $f_{\bar p/p}$ is the PAMELA antiproton/proton flux ratio given in~\cite{Adriani:2010rc}.
Energy bins have been handled in the same way as the positron case,
giving us the upper limit on $\langle v \sigma \rangle$ shown in Fig.~\ref{fig:limits-brem},
 and results using the `minimum' and `maximum' diffusion parameter sets are shown for comparison in Fig.~\ref{fig:limits-comparison}.

%%%%%%%%%%%%%%%%%%%%%%%%%%%%%%%%%%%%%
\subsection{Neutrinos}
\label{sec:neutrinos}
%%%%%%%%%%%%%%%%%%%%%%%%%%%%%%%%%%%%%

Following Ref.~\cite{yhba}, we examine the neutrino flux averaged over the entire sky. The dominant contribution to the DM-signal will be from Galactic annihilations, and we ignore the subdominant cosmic annihilation signal.
We make the approximation that 1/3 of all produced neutrinos will be observed as muon neutrinos,

\begin{equation}
\frac{dN_{\nu_\mu}}{dE}=\frac{1}{3} \frac{dN_\nu}{dE}.
\end{equation}
Note that the neutrino spectrum in Fig.~\ref{fig:spectra} includes primary electron neutrinos,
and all flavors of secondary neutrinos.
We then compare the $\nu_\mu+\bar\nu_\mu$ signal from DM annihilation with the atmospheric $\nu_\mu+\bar\nu_\mu$ flux, using the same technique as Ref.~\cite{yhba}. 
The calculation of the Galactic neutrino signal proceeds the same as for gamma rays, using Eq.~\ref{gamma-flux1} with the neutrino-annihilation spectrum in place of the gamma-ray spectrum.

We average $\mathcal{J}(\psi)$ over the entire sky, and find $\mathcal{J}_{\rm av}\simeq 5$ for the NFW profile. The signal is then compared with the  background flux, integrating each over
an energy bin width of $\Delta \log_{10} E =0.3$.
As expected, the resulting upper limit on the cross section
is significantly weaker than those calculated using the other annihilation products
considered, and is only visible in the high-mass region of Fig.~\ref{fig:limits-brem}.
 Accordingly, the assumptions made in this subsection concerning neutrino flavors are moot.

As an alternative technique, one could calculate the flux of upward-going muons through the Earth induced by neutrinos from DM annihilation, and compare this with limits from the Super-Kamiokande (SuperK) experiment \cite{Desai:2004pq}.
Reference~\cite{Erkoca:2010qx}
 converts the SuperK limits into an upper limit on the annihilation cross section to a neutrino-antineutrino pair. Comparing this with limits on the same channel from Ref.~\cite{yhba}, whose technique we follow,
suggests that the limits presented in this work would not be substantially strengthened by the SuperK data set, certainly not to the point where the neutrino constraints on $\langle v \sigma \rangle_{\rm Brem}$ were competitive with any of the stronger bounds.

%%%%%%%%%%%%%%%%%%%%%%%%%%%%%%%%%%%%%
\section{Discussion}
\label{sec:discussion}
%%%%%%%%%%%%%%%%%%%%%%%%%%%%%%%%%%%%%

%%%%%%%%%%%%%%%%%%%%%%%%%%%%%%%%%%%%%
\begin{figure*}[ht]
\includegraphics[width=0.94\textwidth]{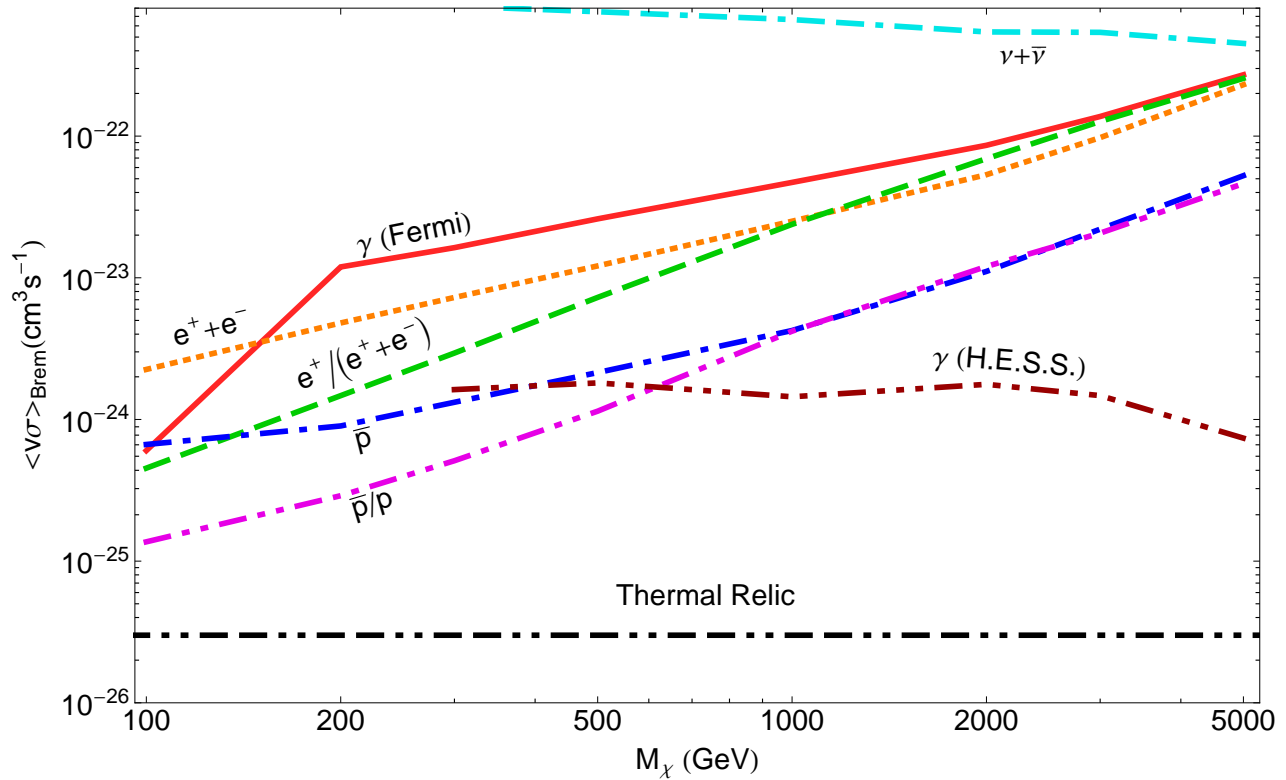}
\caption{Upper limits on $\langle v \sigma \rangle_{\rm Brem}$ using
  the `med' diffusion parameter set.  Shown are constraints based on
  the Fermi EGB (solid, red), 
  $e^++e^-$ flux (dots, orange),
  $e^+/(e^++e^-)$ ratio (dashes, green), 
  $\bar p$ flux (dot-dashes, blue), 
  $\bar p/p$ ratio (dot-dot-dashes, magenta),
  H.E.S.S. gamma rays (dot-dot-dot-dashes, maroon),
  and neutrinos (dot-dash-dashes, cyan).    
  Also shown for comparison is the
  expected cross section for thermal relic dark matter, $3\times
  10^{-26} \,\rm{cm}^3 \, /s$ (dot-dot-dash-dashes, black).
\label{fig:limits-brem}}
\end{figure*} 
%%%%%%%%%%%%%%%%%%%%%%%%%%%%%%%%%%%%%

%%%%%%%%%%%%%%%%%%%%%%%%%%%%%%%%%%%%%
\begin{figure*}[ht]
\includegraphics[width=0.48\textwidth]{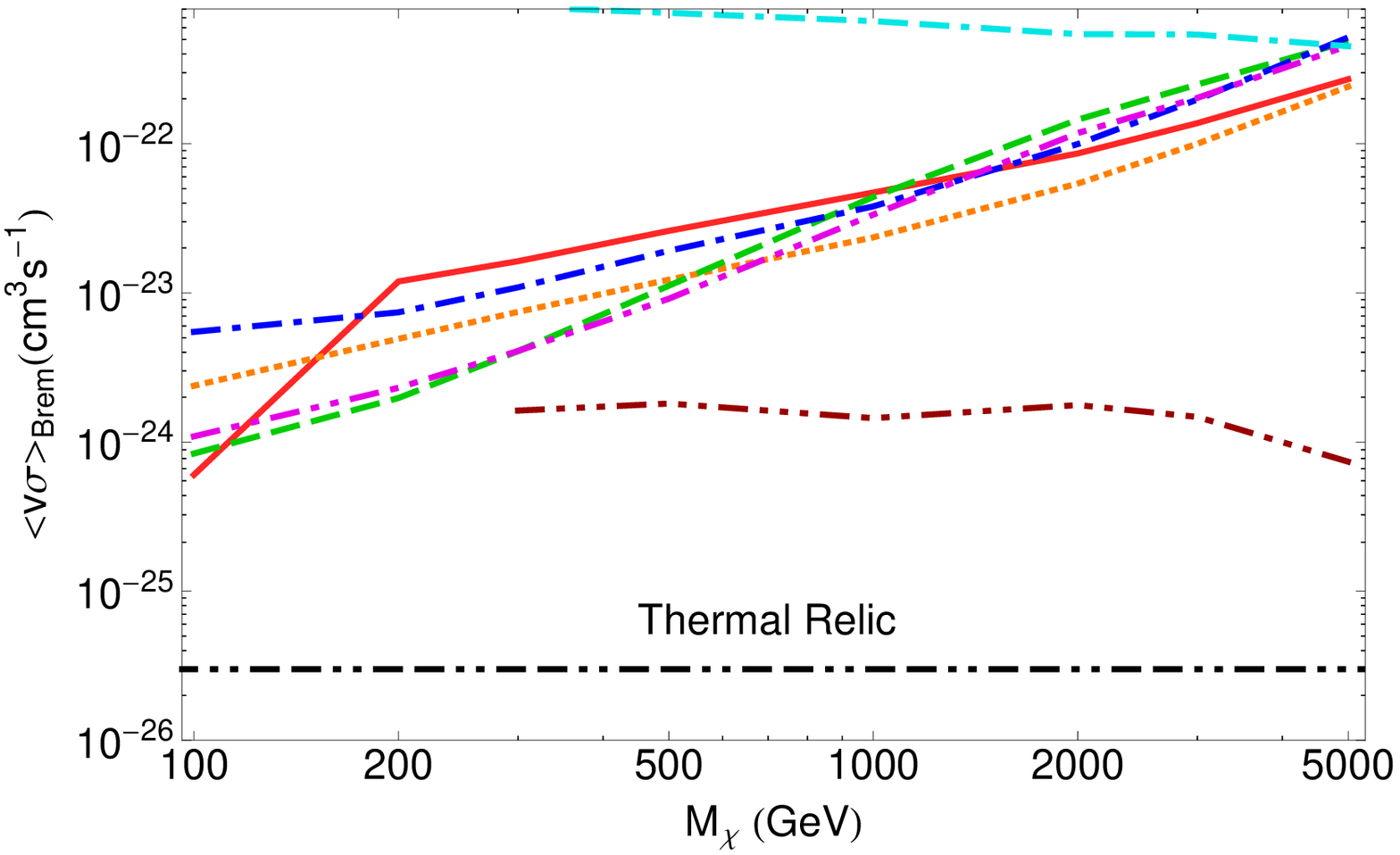} \includegraphics[width=0.48\textwidth]{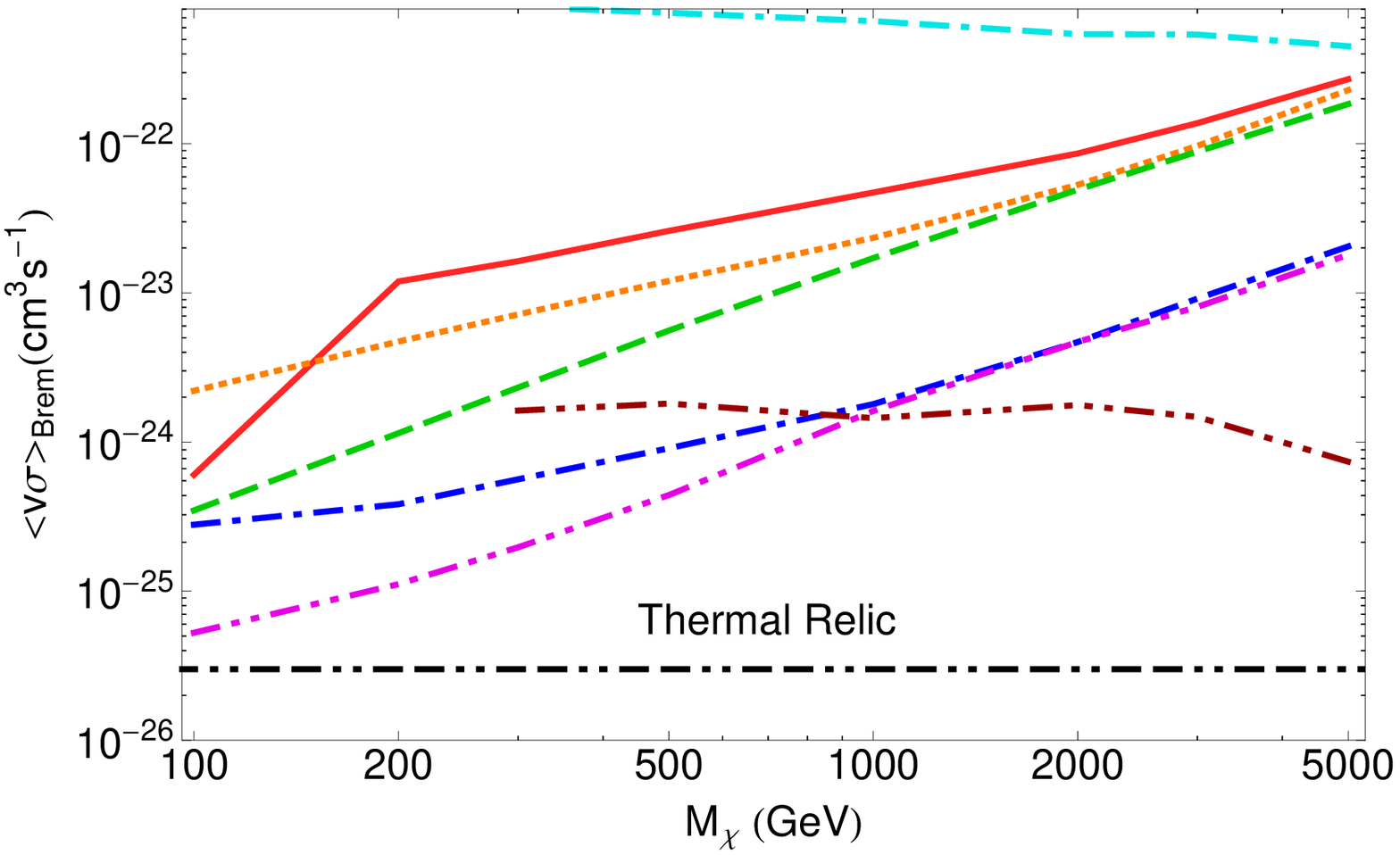}
\caption{As for Fig.~\ref{fig:limits-brem}, using the `min' (left) and `max' (right) diffusion parameter  sets.\label{fig:limits-comparison}}
\end{figure*} 
%%%%%%%%%%%%%%%%%%%%%%%%%%%%%%%%%%%%%

Fig.~\ref{fig:limits-brem} collects our upper limits on the
 bremsstrahlung cross section $\langle v \sigma
\rangle_{\rm Brem}$, as calculated in the previous sections.  The
constraint from the antiproton ratio is stronger than that from the
positron data by a factor of $\sim 5$.
Nature provides a unique value for $\langle v \sigma\rangle$.  
Therefore, if the bremsstrahlung process saturates
the allowed antiproton limit, then the same process produces positrons
at a rate down from the observed value by about a factor of
five.\footnote{Note that we have not compared the spectral shapes of the
  DM signals with those of the observed fluxes, nor tried to {\it fit}
  the data.}  Conversely, if the observed positron fraction were
attributed to the bremsstrahlung process, then the same process would
overproduce antiprotons by about a factor of five.

It is important to note that the observed antiproton flux and ratio
are well reproduced by standard astrophysical processes, leaving
little room for a DM contribution.\footnote{Reference~\cite{Hagiwara} 
notes that a highly anisotropic diffusion
  model, as might be invoked to accommodate galactic winds, may
  suppress the antiproton flux to a value possibly below the PAMELA
  flux.  We do not consider anisotropic diffusion in this work.  } We
have not attempted to model this standard background, so constraints
from antiprotons are likely to be significantly stronger than
presented here.

Annihilation to $\mu^+\mu^-$ or $\tau^+\tau^-$ is not as helicity-suppressed 
as to electrons.  Even so, the helicity factors which
suppress the $s$-wave are $\left(m_\mu/M_\chi\right)^2 \simeq 10^{-7}
\times (M_\chi / 300 \, \rm{GeV})^{-2}$ and
$\left(m_\tau/M_\chi\right)^2 \simeq 3 \times 10^{-5} \times (M_\chi
/300 \, \rm{GeV})^{-2}$, which are comparable to the factor by which
the $p$-wave is velocity suppressed, $v^2 \sim 10^{-6}$.  Since
 bremsstrahlung overcomes both suppressions, it can also be
important for annihilation to muons and taus.  (And, of course, any
helicity suppression is especially stringent for annihilation to
$\nu_\mu$ and $\nu_\tau$, as $m_\nu \simeq 0$.)  In the case of
relatively light DM annihilating to taus, the helicity
suppression is not as pronounced.  
Furthermore, $W$/$Z$~bremsstrahlung may be
kinematically forbidden.  In any case, we note that annihilation to
$\tau$ can never be purely leptophilic, as the $\tau$ has significant
hadronic decay modes.

Note that in the model we consider, 
 emission of photons or massive
gauge bosons can always lift helicity suppression,
however the effect will be much greater in magnitude when
 the DM
and scalar exchange particles are nearly degenerate in mass
(such as the co-annihilation region of mSUGRA).  In the present work
we consider such a region of parameter space.  Specifically, we
present results for $\mu=(M_\eta/M_\chi)^2=1.2$, though our
conclusions remain valid for any value of $\mu$ where the
bremsstrahlung processes dominate the 2 to 2 body processes.  As can
be seen in Ref.~\cite{EWBrem2}, the rates for these two processes
become comparable at $\mu\sim10$.

Consider now scenarios where DM annihilation to a lepton pair
is {\it not} helicity-suppressed. As examples, one may have Majorana
DM annihilating via an exchange of a pseudoscalar or scalar
(the latter is still velocity-suppressed at the DM vertex) or
Dirac DM annihilating via the exchange of a vector, or one
may have scalar DM annihilations. In these cases, there will
still be a signal from electroweak bremsstrahlung emission
\cite{KSS09,BDJW,Ciafaloni:2010b}, although it will no longer be the
dominant channel.  Even so, the $W/Z$ decay products can still lead to
restrictive constraints.
Reference~\cite{KSS09} considered an example (exchange of a scalar) where
EW bremsstrahlung makes only a subdominant contribution to the total
DM annihilation rate.  In this model, the main contribution
to the annihilation rate comes from the 2-body annihilation channels,
thus the monoenergetic $e^\pm$ and neutrino fluxes dwarf the gauge
boson fragmentation products.  Nonetheless, Ref.~\cite{KSS09} found
that the antiproton data still provide the most stringent cross
section constraints for certain parameters.  Note however, that the
models of \cite{KSS09,BDJW} explicitly break gauge invariance.  A
detailed, model-independent, treatment of weak corrections may be
found in Ref.~\cite{Ciafaloni:2010b}.

It should be noted that the results presented here are not due to an
exhaustive survey of all possible DM profiles and parameters.
Uncertainties arise from the various choices made in order to present
illustrative results.  In most cases, we have made conservative
choices for these parameters such that alterations to these selections
should strengthen the results.
In calculating the flux of protons, antiprotons, electrons and positrons, all the astrophysical
parameters are encoded into a numerically-fit function
\cite{Cirelli:2008id} with propagation parameters which are consistent
with a `median' flux \cite{Donato:2003}.  However, by assuming
alternate parameters, e.g. from the `max' or `min' flux scenarios, our
results may be strengthened or weakened by up to an order of
magnitude, as shown in Fig.~\ref{fig:limits-comparison}.
Our conclusions hold in all cases considered, but for the extreme
choice of `min' diffusion parameter set, where the $e^+/(e^++e^-)$
limits become comparable to those for $\bar{p}/p$ and the $e^++e^-$
limits become comparable to those for gamma rays.

Obviously, the choice of profile can have a large effect on the
parameter ranges, and we have adopted the NFW profile with $\rho_\odot
=$ 0.39 GeV/$\textrm{cm}^{-3}$ throughout this work.  If one considers
nonspherical profiles or dark discs then the uncertainty in the value
of the local DM density may be expanded to accommodate a
value between 0.2 GeV/$\textrm{cm}^{-3}$ and 0.7
GeV/$\textrm{cm}^{-3}$\cite{Weber:2009pt}.  
Note, however, that changes to the DM profile would move all the
predicted fluxes, and thus the corresponding cross section
constraints, in the same direction.
For the calculation of the extragalactic fluxes, the cosmic source
clustering factor ($f(z)$ of Eq.(\ref{cosmicflux})) can vary by an
order of magnitude depending on the profile and inclusion or exclusion
of subhalos \cite{yhba} (for example, in the case of the Moore profile
\cite{Moore:1999}, our choice for normalization could be increased by
a factor of 10 \cite{Ando:2005}).  This would lead to tighter
constraints coming from the gamma-ray signals.

We have also neglected the signals produced by inverse Compton
scattering (gamma rays) and synchrotron radiation (radiowaves) of the
electrons and positrons as they propagate in the galaxy (see, e.g.,
Ref.~\cite{Cirelli:2010xx,Crocker:2010gy}).  Note, however, that these
effects are properly included in the electron energy-loss formalism we
adopt.  We expect the gamma rays produced directly from the
annihilations to dominate the constraints.  Additional gamma rays from
inverse Compton would only strengthen our results (but make them less
robust).

\section{Conclusions}

If DM is Majorana in nature, then its annihilation to
fermions may be suppressed due to helicity considerations.  However,
when the DM mass is greater than $M_W/2$, both electroweak
and photon bremsstrahlung may lift this suppression, thereby becoming
the dominant channel for DM annihilation.  This permits the
indirect detection of models for which the annihilation cross section
would otherwise be too suppressed to be of interest.
Subsequent decay of the emitted $W$ and $Z$ gauge boson will produce
fluxes of electrons, positrons, neutrinos, hadrons, and gamma rays.
The aim of the present work has been to study the spectra of these
particles as a tool for indirect detection of DM.  By
comparing these fluxes to cosmic ray data we have been able to
constrain the DM annihilation cross section in such models.
From these constraints, we find that the observational data pertaining
to the flux of antiprotons combined with those of positrons make it
difficult for helicity-suppressed $2\rarr 2$ 
leptophilic DM annihilation to be the source of
the recently detected cosmic ray anomalies.
For these models, the bremsstrahlung processes dominate.
The primary culprit is the
hadronization of the gauge bosons, which leads to a significant
antiproton flux.  This result highlights the difficulty of producing
lepton-only final states even in a model expressly designed for just
such a purpose with $2\rarr 2$ annihilation.

%%%%%%%%%%%%%%%%%%%%%%%%%%%%%%%%%%%%%
\section*{Acknowledgements}
We thank Sheldon Campbell and Yudi Santoso for helpful discussions.
NFB was supported by the Australian Research Council, TDJ was supported 
by the Commonwealth of Australia, JBD was supported in part by the 
Arizona State University Cosmology Initiative, and TJW was supported in 
part by U.S.~Department of Energy Grant No. DE--FG05--85ER40226.
%%%%%%%%%%%%%%%%%%%%%%%%%%%%%%%%%%%%%

\end{document}